\documentclass[15pt]{article}

\usepackage{arxiv}

\usepackage[utf8]{inputenc} 
\usepackage[T1]{fontenc}    
\usepackage{hyperref}       
\usepackage{url}            
\usepackage{booktabs}       
\usepackage{amsfonts}       
\usepackage{nicefrac}       
\usepackage{microtype}      
\usepackage{lipsum}
\usepackage{amsmath}
\usepackage[labelfont=bf]{caption}
\usepackage{siunitx}
\usepackage{bm}
\usepackage{color, colortbl} 
\usepackage{multirow}
\usepackage{graphicx}
\usepackage{setspace}
\usepackage[superscript]{cite}
\usepackage{float}
\usepackage{siunitx}
\usepackage{enumerate}

\begin{document}
\captionsetup[figure]{labelfont={bf},name={Fig.},labelsep=period}
\vspace*{0.35in}


\begin{flushleft}
{\Large
\textbf\newline{\bf{First and Second Order Raman Spectroscopy of Monoclinic \bm{$\beta-\mathrm{Ga}_2\mathrm{O}_{3}$}} }
}
\newline
\\
Benjamin M. Janzen \textsuperscript{1},
Roland Gillen \textsuperscript{2},
Zbigniew Galazka \textsuperscript{3},
Janina Maultzsch \textsuperscript{2},
and Markus R. Wagner \textsuperscript{1,*}
\line(1,0){470}
\\
\vspace{.3cm}
\bf{1:} \normalfont Technische Universität Berlin, Institute of Solid State Physics, Hardenbergstraße 36, 10623 Berlin, Germany.
\\
\bf{2:} \normalfont Chair of Experimental Physics, Friedrich-Alexander Universität Erlangen-Nürnberg, Staudtstraße 7, 91058 Erlangen, Germany.
\\
\bf{3:} \normalfont Leibniz-Institut für Kristallzüchtung, Max-Born-Str. 2, 12489 Berlin, Germany.
\\
\vspace{.5cm}
* markus.wagner@physik.tu-berlin.de
\line(1,0){470}

\end{flushleft}

\begin{abstract}
We employ a combined experimental-theoretical study of the first- and second-order Raman modes of monoclinic \mbox{$\beta$-Ga$_{2}$O$_{3}$}. The investigated materials is of particular interest due to its deep-UV bandgap paired with a high critical field strength, offering promising applications in power-electronics. 
A crucial prerequisite for the future development of \mbox{Ga$_{2}$O$_{3}$}-based devices is a detailed understanding of the lattice dynamics as they are important for the elasticity (through acoustic phonons), thermal conductivity (through the heat transferred by phonons), the temperature-dependence of the bandgap (impacted by electron-phonon coupling) or the free carrier transport (via phonon scattering).
Polarized micro-Raman spectroscopy measurements on the (010) and ($\bar{2}01$) planes enable the determination of the phonon frequencies of all 15 first-order and more than 40 second-order Raman modes. The experimental results are correlated with calculations of the mode frequencies, phonon dispersion relation and phonon density of states using density functional perturbation theory (DFPT). By applying a group-theoretical analysis, we are able to distinguish between overtones and combinational modes and identify the high symmetry points in the Brillouin zone which contribute to the observed second order modes. Based on these information, we demonstrate the simultaneous determination of Raman-, IR-, and acoustic phonons in \mbox{$\beta$-Ga$_{2}$O$_{3}$} by second-order Raman spectroscopy.

 \end{abstract}

\section{Introduction}

A material's crystal structure and lattice dynamics strongly affect crucial material properties including elasticity (through acoustic phonons), thermal conductivity (through the heat transferred by phonons), the temperature-dependence of the bandgap (impacted by electron-phonon coupling) or the free carrier transport (via phonon scattering). A powerful tool to investigate a material's vibrational properties is the employment of Raman spectroscopy, which provides insights into the Raman-active phonon modes. 
\newline The majority of the Ga$_{2}$O$_{3}$-related Raman studies have focused on the thermodynamically most stable monoclinic $\beta$ polymorph\cite{Kranert,Dohy,Machon,Liu,Onuma,Janzen2021,Zhang2021_beta,Zhang2021_T-dpendent,Yao2019}, although a set of polarized Raman spectra of the orthorhombic $\kappa$ (also referred to as $\epsilon$)\cite{Cora} phase was recently reported\cite{Janzen2021b}.
The first-order Raman spectra of $\beta$-Ga$_{2}$O$_{3}$ single crystals were already demonstrated in 1982 by Dohy et al.\cite{Dohy}.
The authors employed temperature-dependent Raman spectroscopy to study the influence of lattice expansion. In conjunction with valence force field calculations they identified three categories of Raman-active phonons with respect to the motions of Ga and O atoms. A following work by Machon et al.\cite{Machon} carried out Raman spectroscopy in diamond anvil cells, revealing a transition from the $\beta$- to the $\alpha$-phase under high pressure. The first set of polarized Raman spectra of $\beta$-Ga$_{2}$O$_{3}$ enabling to distinguish all 15 Raman modes of different vibrational symmetries was added by Onuma et al.\cite{Onuma} in 2014. Fiedler et al.\cite{Fiedler2020} investigated the effects of high Si- or Sn-doping concentrations on the first-order Raman spectra of \mbox{$\beta-\textrm{Ga}_{2}\textrm{O}_{3}$} and observed a number of additional Raman peaks, whose excitations were assigned to (i) electronic excitations involving the impurity band formed by effective-mass-like hydrogenic shallow donors, (ii) non-hydrogenic donors or the (iii) phonon-plasmon coupling associated with infrared-active modes. Yao et al.\cite{Yao2019} employed Raman spectrocopy and Raman spectral mapping as a non-destructive method to evaluate the crystallinity and uniformity of single-crystal \mbox{$\beta-\textrm{Ga}_{2}\textrm{O}_{3}$} grown by the edge-defined film-fed growth (EFG) method. Similarly, in a pair of studies Zhang et al. employed confocal\cite{Zhang2021_beta} and temperature-dependent\cite{Zhang2021_T-dpendent} Raman spectroscopy to investigate Si-, Mg-, Fe- and Sn-doped EFG-grown \mbox{$\beta$-Ga$_{2}$O$_{3}$}, proving superior crystallinity and high uniformity of both the un- and ion-doped Ga$_{2}$O$_{3}$ substrates. The authors revealed frequency shifts and intensity variations of the intrinsic Raman modes upon ions doping, which were ascribed to the substitution of dopants on Ga lattice sites.\cite{Zhang2021_beta} Moreover, they revealed that both the Raman frequencies and full widths at half-maximum (FWHM) of the individual Raman modes scaled linearly with temperature in a range between \mbox{77–297 K} for both the doped and undoped samples.\cite{Zhang2021_T-dpendent} In a subsequent study, Seyidov et al.\cite{Seyidov2022} applied Raman spectroscopy of Mg- and Si-doped bulk \mbox{$\beta$-Ga$_{2}$O$_{3}$} crystals grown by the Czochralski method. Exclusively for the Mg-doped sample, the authors observed an additional Raman peak at \mbox{5150 cm$^{-1}$,} which was attributed to resonant electronic Raman scattering (ERS) produced by Ir$^{4+}$ ions.
\newline The impact of the individual lattice sites on the Raman spectra was further investigated by Janzen et al.\cite{Janzen2021} who studied \mbox{$\beta$-Ga$_{2}$O$_{3}$} in two different oxygen isotope compositions ($^{16}$O,$^{18}$O). Quantifying the mode frequency shifts of all Raman modes observed for the $^{18}$O with respect to the $^{16}$O spectra enabled to identify the (i) atomistic origin of all modes (Ga-Ga, Ga-O or O-O) as well as the (ii) Raman modes that were dominated by the different, inequivalent O- or Ga-atoms of the unit cell.
Moreover, Kranert et al.\cite{Kranert} experimentally determined the Raman tensor elements in the framework of a modified Raman tensor formalism proposed in their preceding study\cite{Kranert_Grundlagen}. Aside from the Raman-active phonons, IR-active phonons constitute another important class of quasi-particles and were treated theoretically in a number of publications.\cite{Liu,Onuma2016,Schubert,Mengle2019,Dohy} Experimental access to infrared-active TO-phonons was first provided by Dohy et al.\cite{Dohy} by employing IR transmittance spectroscopy. The results were confirmed and extended by Villora et al.\cite{Villora2002} who applied IR reflectance spectroscopy. The two consecutive works of Onuma et al.\cite{Onuma2016} and Schubert et al.\cite{Schubert} added the LO-phonons by using IR ellipsometry. 
\newline 
Nevertheless, the simultaneous detection of IR- and Raman-active phonons is not possible using ellipsometry or first-order Raman spectroscopy. One possibility to overcome this issue is the employment of second-order Raman spectroscopy (also referred to as two-phonon Raman scattering), in which two single phonons interact with an incident photon to produce Raman overtones or sum (difference) frequency combinations. While analogously to first-order Raman scattering energy and momentum conversation must apply for this many-particle interaction, the momentum conservation condition permits the participation of phonons with arbitrary wave vectors. Consequently, the restriction to $\Gamma$-point phonons as is the case in first-order Raman experiments is lifted, allowing not only to probe Raman-active phonons from across the Brillouin zone, but also to study the IR-active phonons previously inaccessible in first-order Raman scattering.
Whereas second-order Raman scattering has been thoroughly investigated for cubic\cite{Birman1962,Kleinman1965,Parker1967,Weinstein1973,Weinstein1973_GaP,Mishra2001} and hexagonal\cite{Calleja1977,Murugkar1995,Siegle1997,Davydov1998,Haboeck2003,Gao2007,Hu2019} crystals, studies on monoclinic systems are scarce.
\newline 
Using high-resolution polarized micro-Raman spectroscopy to probe bulk \mbox{$\beta$-Ga$_{2}$O$_{3}$} single crystals grown by the Czochralski method\cite{Galazka2010,Galazka2014,Galazka2017}, we provide a comprehensive study of first- and second-order Raman spectra. Phonon modes of different vibrational symmetries are separated by selected polarization configurations of the incident and scattered light. Aside from determining the site symmetry groups of the high-symmetry points within the Brillouin zone, we perform a group-theoretical analysis to derive second-order Raman selection rules. In combination with second-order selection rules, the calculated phonon dispersion curves as well as the single-phonon density of states (PDOS), the experimental data enable to identify the origins (points within the Brillouin zone where scattered phonons originate from) and nature of second-order events, i.e. whether the observed modes are combinations of single phonons from different dispersion branches or overtones (two phonons from the same branch). Our detailed second-order Raman analysis provides the simultaneous experimental investigation of the IR- and Raman-active phonons from across the Brillouin zone.

\section{Materials and methods}

Crystal samples of size 5x5x0.5 mm$^{3}$ used in the present study were prepared from undoped, two inch diameter single crystal grown by the Czochralski method utilizing an Ir crucible and oxidizing growth atmosphere, as described in detail elsewhere\cite{Galazka2017,Galazka2021}. 
The preparation included different surface orientations of (100), (010), (001), and $(\bar{2}01)$ and chemical-mechanical polishing (CMP). The samples were semiconducting with the free electron concentration and electron mobility of \mbox{$3.4\cdot 10^{17}$ cm$^{-3}$} and \mbox{118 cm$^{2}$V$^{-1}$s$^{-1}$} according to Hall effect measurements\cite{Irmscher2011}.

Raman scattering at room temperature (293 K) was induced by a 532.16 nm frequency stabilized single longitudinal mode diode-pumped solid-state (DPSS) laser (Laser Quantum Torus 532) on a LabRAM HR 800 spectrometer (Horiba Jobin-Yvon). The laser beam was focused onto the sample using a 50x Olympus objective with a numerical aperture (NA) of 0.75, with the scattered light being collected in backscattering geometry. Backreflected and elastically scattered light (Rayleigh component) was filtered using an ultra low frequency filter (ULF) unit and then spectrally-dispersed by a monochromator with a grating of 600 lines/mm. The light was detected by a charge-coupled device (CCD).
The sample was placed beneath the objective with a respective surface's normal parallel to the direction of light propagation. A $\lambda/2$ wave plate in the excitation was set at 0$^{\circ}$ or 45$^{\circ}$ to polarize the incident light parallel or crossed with respect to the scattered light, which was selected using a fixed polarizer in the detection. 
Prior to each measurement, the Raman spectrometer was calibrated using the spectral lines of a neon spectral lamp.

The theoretical results were computed using density functional perturbation theory (DFPT) as implemented into the Quantum Espresso suite~\cite{qe}. The local-density approximation (LDA) to the exchange-correlation interaction has been shown ~\cite{Janzen2021,Janzen2021b} to yield a good description of the vibrational spectra of Ga$_2$O$_3$ and was thus used for all computations. We also did calculations using the popular PBEsol functions, which yielded a convincing qualitative agreement with the LDA results, albeit with an underestimation of the computed phonon frequencies.
We fully optimized the atomic positions and cell parameters until the residual forces between atoms and the cell stress were smaller than 0.001\,eV/\AA\space and 0.01\,GPa, respectively. The threshold for the total energy was set to 10$^{-15}$\,Ry, which ensured tightly converged interatomic forces and electronic density distributions. The Ga($3d$,$4s$,$4p$) and the O($2s$,$2p$) states were treated as valence electrons using multi-projector normconserving pseudopotentials from the PseudoDojo library~\cite{pseudodojo} with a cutoff of 180\,Ry. All reciprocal space integrations of the electronic structure were performed on a $\Gamma$-centered Monkhorst-Pack grid of 8x8x8 $k$-points in the Brillouin zone. We then computed the phonon modes and dynamical matrices for phonon momenta on a regular grid of 4x4x4 $q$-points. Based on these results, the phonon band structure and corresponding phonon density-of-states (phDOS) were obtained through Fourier interpolation along a chosen path between high-symmetry points and onto a denser grid of 50x50x50 q-points, respectively.

\section{Results and discussion}

\subsection{First-order Raman scattering}
\begin{figure}[!htb]
\centering 
        \includegraphics[width=160mm]{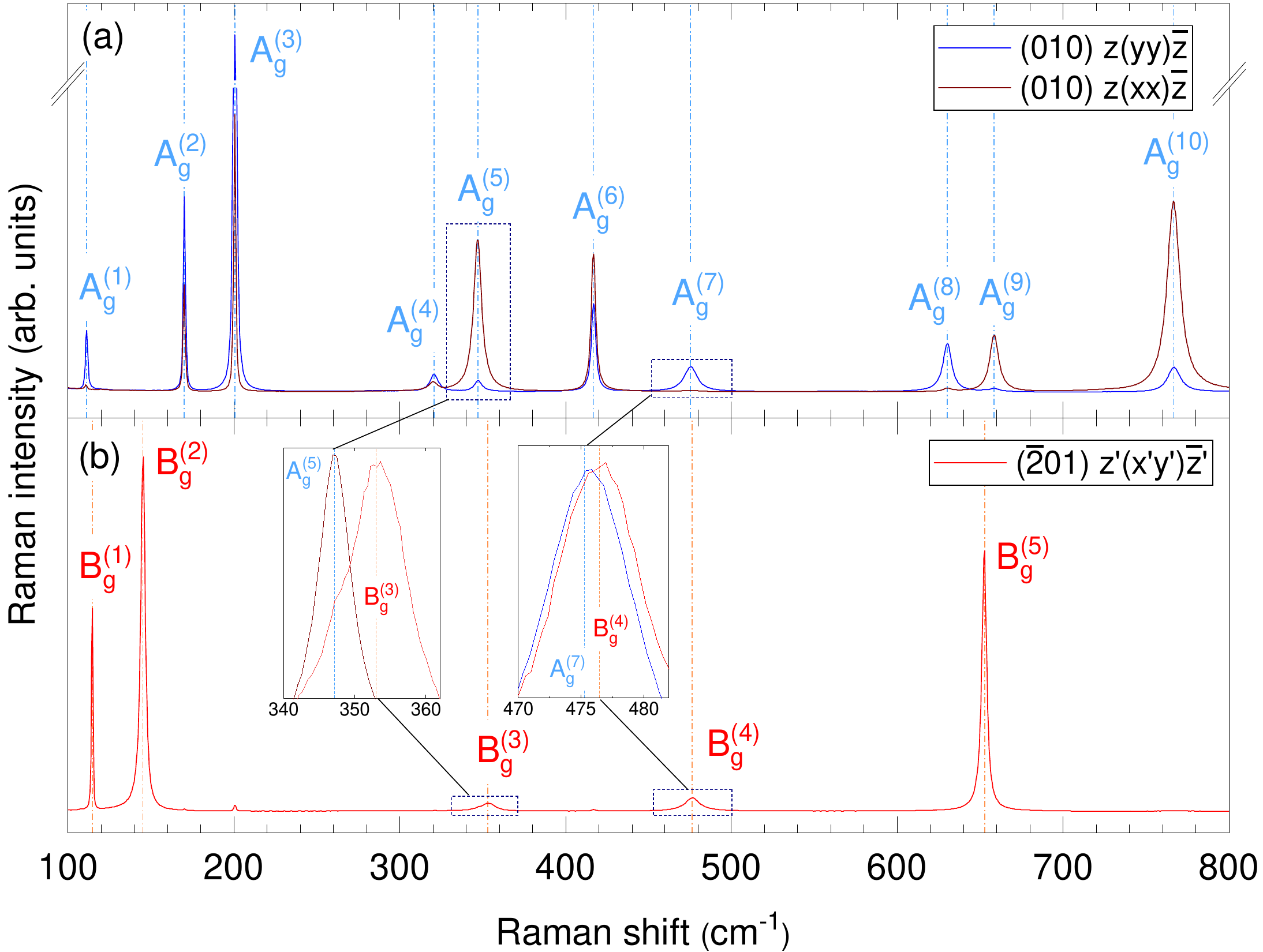}
    \caption[Raman spectra of the investigated $\beta$-Ga$_{2}$O$_{3}$ single crystal]{ Raman spectra of the investigated $\beta$-Ga$_{2}$O$_{3}$ single crystal. $A_{g}$-modes are obtained by irradiating the (010) plane in parallel polarization, with Cartesian coordinates $x$, $y$ and $z$ chosen such that the $z$-axis aligns with the [010] direction, while $y\parallel [100]$. $B_{g}$-modes are acquired by excitation of the ($\bar{2}01$) plane in crossed polarization. Here, $z'$ points into the direction of the surface normal, with $x'\parallel[010]$ and $y'\parallel[102]$. The insets resolve the closely-matching $A_{g}^{(5)}/B_{g}^{(3)}$ or $A_{g}^{(7)}/B_{g}^{(4)}$ modes with magnified scaling. \label{Figure: Raman spectra of the investigated beta-Ga2O3 plates}}
\end{figure}

The first-order Raman-active phonon modes of $\beta-$Ga$_{2}$O$_{3}$ are acquired in polarized micro-Raman measurements, the results of which are depicted in \mbox{Fig. \ref{Figure: Raman spectra of the investigated beta-Ga2O3 plates}.}

The monoclinic unit cell (space group: $C_{2\mathrm{h}}^{3}$; $C2/m$)\cite{Furthmueller,Yoshioka,Cora} is composed of ten atoms: four Ga-atoms (two each in tetrahedral and octahedral coordination) and six O-atoms in between.\cite{Furthmueller} This implies a total of 30 phonons within the first Brillouin zone, whose irreducible representation at the $\Gamma$-point is given by\cite{Dohy,Onuma}
\begin{equation}
    \Gamma=10\ A_{g}+5\ B_{g}+5\ A_{u}+10\ B_{u}=\Gamma_{\textrm{aco}}+\Gamma_{\textrm{opt}},
\end{equation}

where \mbox{$\Gamma_{\textrm{aco}}=A_{u}+2\ B_{u}$} or \mbox{$\Gamma_{\textrm{opt}}=10\ A_{g}+5\ B_{g}+4\ A_{u}+8\ B_{u}$} refer to the acoustic or optical phonons, respectively. Among the optical phonons, modes with even parity (index $g$) are Raman-active and can be probed in first-order Raman measurements, whereas modes of odd parity (index $u$) are IR-active and hence not detectable using first-order Raman spectroscopy. 
\newline The Raman-active first-order phonon modes of different vibrational symmetries are separated by application of specific polarization geometries as predicted by first-order Raman selection rules (summarized in \mbox{Table \ref{Table: Selection rules_beta-polymorph}).}

\definecolor{Gray}{gray}{0.9}
\begin{table}[ht]
\renewcommand{\arraystretch}{1.4}
\caption[Raman selection rules of monoclinic $\beta$-Ga$_{2}$O$_{3}$]{First-order Raman selection rules for monoclinic crystals in typical backscattering geometries. Polarization configurations are described using the Porto notation. Coordinates $x$, $y$, $z$ are chosen such that $y$ and $z$ align with the [100] or [010] directions, while the $x$-axis is tilted an angle of $13.8^{\circ}$ against the $[001]$ direction. The $z'$ direction points into the direction of the ($\bar{2}01$) plane's surface normal, with $x'\parallel [010]$ and $y'\parallel$[102].
} 
\centering
\begin{tabular}{|c|c|c|}\rowcolor{Gray}\hline  
		
		\textbf{Plane} & \textbf{Polarization} \bm{$\vec{k}_{i}(\vec{e}_{i}\vec{e}_{s})\vec{k}_{s}$}& \textbf{Allowed Raman modes} \\ \hline
		$b$/(010)&$z(yy)\bar{z},z(xx)\bar{z}$&$A_{g}$  \\ \hline 
		($\bar{2}$01)&$z'(x'x')\bar{z}',z'(y'y')\bar{z}'$&$A_{g}$  \\ \hline 
		($\bar{2}$01)&$z'(x'y')\bar{z}'$&$B_{g}$  \\ \hline 
	
		\end{tabular}
		\setlength\belowcaptionskip{5pt}\\
		\label{Table: Selection rules_beta-polymorph}
		\end{table}   

Using the Porto notation, the scattering geometries applied on the (010) plane in parallel polarization \mbox{(Fig. \ref{Figure: Raman spectra of the investigated beta-Ga2O3 plates}a)} can be written as (i) $z(yy)\bar{z}$ (blue) or (ii) $z(xx)\bar{z}$ (brown), where $y$ and $z$ correspond to the [100] or [010] directions, while the $x$-axis is tilted an angle of $13.8^{\circ}$ against the $[001]$ direction (cf. Fig. 1 in reference \cite{Janzen2021}). Selection rules predict the presence of $A_{g}$ and absence of $B_{g}$ modes in these configurations. 
Conversely, solely $B_{g}$ modes are available when exciting the ($\bar{2}01$) plane and selecting a crossed polarization configuration $z'(x'y')\bar{z}'$ \mbox{(Fig. \ref{Figure: Raman spectra of the investigated beta-Ga2O3 plates}b),} with $x'\parallel [010]$ and $y'\parallel$[102].
With regard to the measurements on the (010) plane our data reveal that the (i) $A^{1}_{g}$, $A^{2}_{g}$, $A^{3}_{g}$, $A^{4}_{g}$, $A^{7}_{g}$ and $A^{8}_{g}$ or (ii) $A^{5}_{g}$, $A^{6}_{g}$, $A^{9}_{g}$ and $A^{10}_{g}$ modes have maximum intensity in the respective geometries, indicating a 90$^{\circ}$ phase shift between the two races of modes previously observed in angular-resolved Raman scans\cite{Kranert}.
\newline
The spectral positions of the Raman modes in \mbox{Fig. \ref{Figure: Raman spectra of the investigated beta-Ga2O3 plates}} are determined by fitting Lorentzian lineshape functions, with the peak positions listed in \mbox{Table \ref{Table: Spectral positions of single crystal beta-Ga2O3modes}}. Our results show good agreement with previous experimental and theoretical data. 

\renewcommand{\arraystretch}{1.3}
\definecolor{Gray}{gray}{0.9}
\begin{table}[!htp]
\caption[Spectral positions of the $\beta-\mathrm{Ga}_{2}\mathrm{O}_{3}$ first-order Raman peaks ]{Spectral positions of the $\beta-\mathrm{Ga}_{2}\mathrm{O}_{3}$ first-order Raman peaks in Fig. \ref{Figure: Raman spectra of the investigated beta-Ga2O3 plates}, given in cm$^{-1}$. Peak positions were determined by fitting Lorentzian lineshape functions. The data are compared with the results of DFPT-LDA calculations and previous experimental as well as theoretical results. The results obtained in this work are denoted by $\dagger$. Where available, brackets () specify the applied exchange correlation functional. A more likely assignment for two peaks from \cite{Machon} suggested by \cite{Kranert} is indicated by “$^{*}$”. “-” denotes modes which were not observed. \label{Table: Spectral positions of single crystal beta-Ga2O3modes}}
\centering
\resizebox{\textwidth}{!}{%
\begin{tabular}{|c|c|c|c|c|c|c|c|c|c|c|c|c|}
\hline
\rowcolor{Gray}  & \multicolumn{6}{c|}{Experiment} & \multicolumn{6}{c|}{Theory} \\ \cline{2-11} 
\rowcolor{Gray}  Phonon&  & &  &   &   &    & &    &    & &    & (Perdew-Burke-  \\ 
\rowcolor{Gray}  mode&  $\dagger$ & \cite{Janzen2021} &  \cite{Kranert}  & \cite{Dohy}  &  \cite{Machon}  & \cite{Onuma}  &  (LDA) $\dagger$ & (LDA)\cite{Janzen2021} &  (B3LYP)\cite{Kranert}  & \cite{Dohy}&  (LDA)\cite{Machon}   & Ernzerhof)\cite{Liu}   \\ \hline
 $A_{g}^{1}$ &  111.4 & 110.7& 111.0   & 111   &  110.2  & 112  & 106.0 &106.4  &  113.5  &  113  & 104   & 104.7  \\ \hline
 $B_{g}^{1}$&  114.8 &114.3 &  114.8  &   114 &  113.6  & 115  &107.1 & 107.7  &  118.6 &   114 &  113  & 112.1  \\ \hline
$B_{g}^{2}$&  145.5 & 144.9 &  144.8  &  147  &  144.7  &  149 & 146.6& 145.0  &  145.6  &   152 &  149  &  141.3 \\ \hline 
$A_{g}^{2}$ &  170.4 & 169.8 &  169.9  & 169 &  169.2  & 173  & 163.4& 163.1  &  176.4  &   166 &  165  & 163.5  \\ \hline
$A_{g}^{3}$&   200.9 & 200.3 & 200.2  & 199&  200.4  & 205  & 191.6 & 190.5 &  199.1  &   195 &  205  & 202.3  \\ \hline
$A_{g}^{4}$&  320.9  & 320.3&  320.0  &  318  & 318.6  & 322  &314.9 & 314.0  &  318.5  &   308 &  317  &  315.8 \\ \hline
$A_{g}^{5}$ &  347.2  & 347.0 &  346.6  & 346   &  364.4  &  350 &348.5 & 345.0  &  342.5  &  353  &  346  & 339.7  \\ \hline
$B_{g}^{3}$&  353.5 & 353.4 & 353.2  &  353  & -  & 355  &355.2 & 351.4  &  359.2  &   360 &  356  & 348.3  \\ \hline
$A_{g}^{6}$&   417.0 & 416.7 & 416.2  & 415   & 415.7  & 421  & 382.6& 384.2  &  432.0  &  406 &  418  & 420.2  \\ \hline
$A_{g}^{7}$&  475.6 & 475.3 &  474.9  &  475  & -   &  479 &461.7 & 458.9  &  472.8  &  468  &  467  & 459.4  \\ \hline
$B_{g}^{4}$&  476.6 & 475.9 &  474.9  &  475  &  473.5  &  480 &477.6 & 473.3  &  486.1  &  474  &  474  & 472.8  \\ \hline
$A_{g}^{8}$&  630.4 & 630.4 &  630.0  &  628  & 628.7$^{*}$    & 635  & 626.0& 620.3   & 624.4   &  628  &  600  &  607.1 \\ \hline
$B_{g}^{5}$&  652.6 & 652.4 &  652.3  &  651  &  652.5$^{*}$  &  659 & 647.3& 644.4  & 653.9   &  644  &  626  &  627.1 \\ \hline
$A_{g}^{9}$&  658.5 & 659.0 &   658.3 &  657  &  -$^{*}$  & 663  &652.4 & 648.5  &   655.8 &  654  &  637  & 656.1  \\ \hline
$A_{g}^{10}$&  766.9 & 767.3 &  766.7  &  763  & 763.9   & 772  &754.4 & 751.5  &  767.0  &  760  & 732   & 757.7  \\ \hline
\end{tabular}
}
\end{table}

The vast majority of experimental works does not distinguish between the $A_{g}^{7}$ and $B_{g}^{4}$ modes due to their spectral proximity. In fact, only two previous studies\cite{Onuma,Janzen2021} have reported their individual frequencies. A reliable differentiation between these modes requires sufficiently strong suppression of either $A_{g}$ or $B_{g}$ modes by choice of scattering geometry and polarization. In the present study, this can easily be demonstrated by the intensity of the commonly most intense $A_{g}^{3}$ mode in the parallel-polarized scattering geometries on the (010) plane in \mbox{Fig. \ref{Figure: Raman spectra of the investigated beta-Ga2O3 plates}a,} which is barely visible in the crossed polarized spectra of the ($\bar{2}01$) plane in \mbox{Fig. \ref{Figure: Raman spectra of the investigated beta-Ga2O3 plates}b.}  Conversely, the most intense $B_{g}^{2}$ mode in \mbox{Fig. \ref{Figure: Raman spectra of the investigated beta-Ga2O3 plates}b} is completely absent in \mbox{Fig. \ref{Figure: Raman spectra of the investigated beta-Ga2O3 plates}a.}
Consequently, the positions of closely neighboring modes of different symmetries ($A_{g}^{7}$ and $B_{g}^{4}$ or $A_{g}^{5}$ and $B_{g}^{3}$) can be determined with high confidence without interference of the respective others. In detail, we obtained a frequency difference of \mbox{1.0 cm$^{-1}$} between the $A_{g}^{7}$ and $B_{g}^{4}$ modes, which is in line with the previously reported values\cite{Onuma,Janzen2021} ranging between \mbox{0.6 cm$^{-1}$} and \mbox{1.0 cm$^{-1}$}.

\subsection{Second-order Raman scattering}

\begin{figure}[!htb]
\centering 
        \includegraphics[width=165mm]{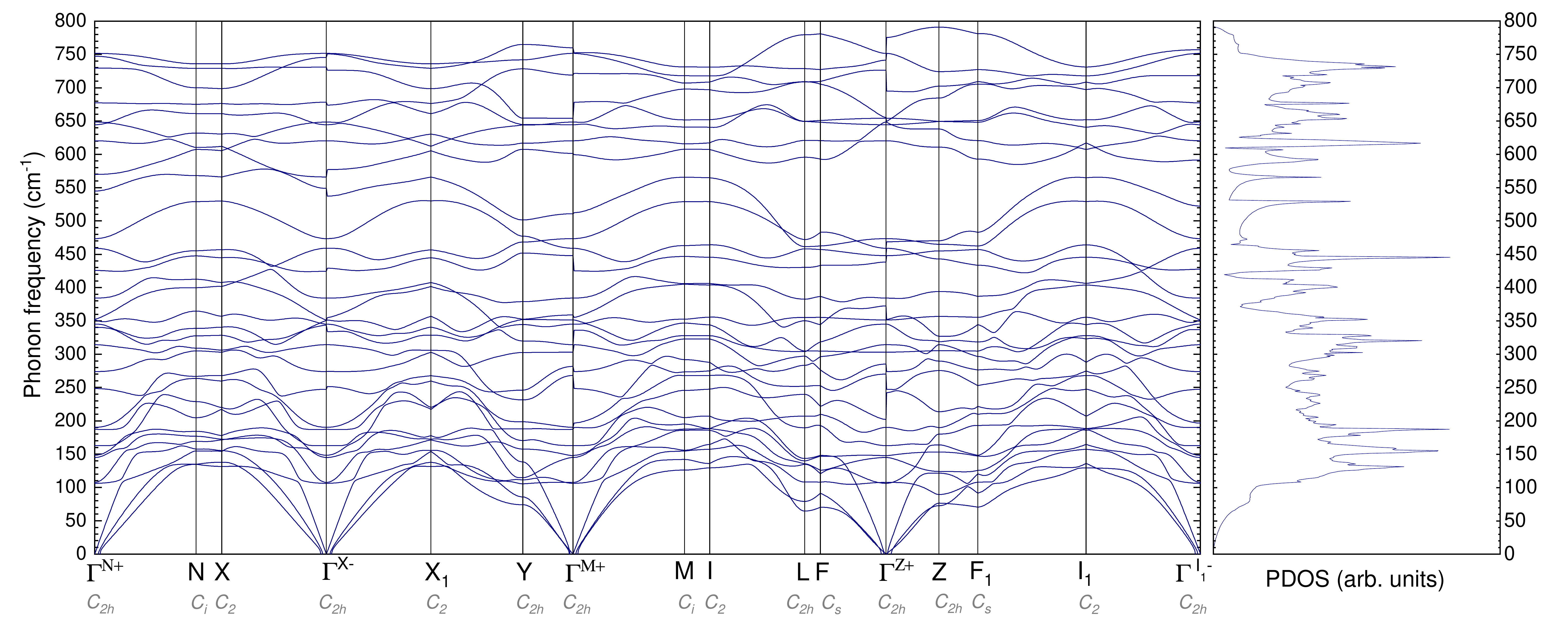}
    \caption[Calculated phonon dispersion and phonon density of states of monoclinic $\beta$-Ga$_{2}$O$_{3}$]{Calculated phonon dispersion (left) and phonon density of states (right) of monoclinic \mbox{$\beta$-Ga$_{2}$O$_{3}$.} The critical points in the monoclinic unit cell along with their corresponding site group symmetries are illustrated at the bottom of the graph. Individual $\Gamma$-points are distinguished by introducing supercripts to label each $\Gamma$-point by the inbound (-) or outbound (+) phonon branch direction, e.g. $\Gamma^{X-}$ or $\Gamma^{X_{1}+}$ as equivalent labels for the $\Gamma$-point between the $X$ and $X_{1}$ points. "PDOS" denotes phonon density of states.   \label{Figure: Phonon dispersion and PDOS}}
\end{figure}

While momentum and energy conservation restrict first-order Raman scattering to the $\Gamma$-point and provides access only to the Raman-active phonon modes, both Raman- and IR- active phonons from across the first Brillouin zone (i.e. with arbitrary wave vectors) participate in two-phonon scattering events. The second-order Raman cross-section and hence the second-order intensity is proportional to the single-phonon density of states, which in turn depends on the phonon dispersion.\cite{Murugkar1995,Siegle1997,Gao2007} A high phonon density of states corresponds to dispersion bands running flat at certain phonon wave vectors $\bm{q}$. Thus, high phonon densities of states within the first Brillouin zone may be observed at critical points, i.e. high-symmetry points where the phonon dispersion becomes flat. The origins of second-order Raman peaks may be discussed based on the phonon dispersion curves and corresponding single phonon density of states (PDOS) in combination with second-order Raman selection rules. The dispersion curves and PDOS are illustrated in \mbox{Fig. \ref{Figure: Phonon dispersion and PDOS}.} For reasons of clarity and simplicity, the phonon branches' symmetries are not indicated, but are available in \mbox{Fig. \ref{Figure: Phonon dispersion and PDOS_mit Labels}} of the Supplementary Material. Polar phonons induce an oscillating macroscopic electric field in the direction of atomic displacements, leading to a splitting of the IR-active $A_{u}$ and $B_{u}$ modes into transverse-optical (TO) and longitudinal-optical (LO) phonons.\cite{Mengle2019,Schubert,Liu} The non-polar $A_{g}$ and $B_{g}$ modes, in turn, exhibit no LO-TO-splitting. The inclusion of the LO-TO-splitting into our calculations leads to a step-like behavior in the phonon dispersion for a number of branches at the $\Gamma$-point, which is fully consistent with previous theoretical studies\cite{Liu,Santia2015,Schubert2020}. 

Since second-order Raman selection rules for monoclinic crystals are presently unavailable, we performed a group-theoretical analysis, the results of which are summarized in \mbox{Table \ref{Table: Selection rules_second order}.} The first column lists the critical points, along with the respective reciprocal space coordinates as determined by a previous work\cite{Peelaers2015}, with the second column providing the corresponding site group symmetries. 

\definecolor{Gray}{gray}{0.9}
\begin{table}[!htp]
\renewcommand{\arraystretch}{1.4}
\centering
\caption[]{Group-theoretical selection rules for two-phonon Raman scattering in monoclinic crystals. "B.Z." or "rsc" denote "Brillouin Zone" or "reciprocal space coordinates". RSC are chosen based on the primitive monoclinic unit cell suggested in reference\cite{Peelaers2015}. The here used reciprocal space coordinates have the following meaning: \mbox{$\Psi=\frac{3}{4}-b^{2}/(4a^{2}\mathrm{sin}^{2}\beta)$,} \mbox{$\Phi=\Psi-(\frac{3}{4}-\Psi)\frac{a}{c}\mathrm{cos}\ \beta$,} \mbox{$\zeta=(2+\frac{a}{c}\mathrm{cos}\beta)/4\ \mathrm{sin}^{2}\beta$} and \mbox{$\eta=\frac{1}{2}-2\zeta\frac{c}{a}\mathrm{cos}\beta$}, with \mbox{$\beta=103.77^{\circ}$} denoting the monoclinic angle\cite{Furthmueller}.} 
\resizebox{\textwidth}{!}{ 
\begin{tabular}{|c|c|c|c|c|}
\hline
\rowcolor{Gray}Point in the  & Site group
 & Irreducible representations for points  &  &  \\
\rowcolor{Gray}B.Z. with rsc & symmetry &  in
the B.Z. and their correlations & Overtones & Combinations  \\ \hline
$\Gamma$ (0,0,0) & $C_{2h}$ & $A_{g}=\Gamma_{1}$  & $[\Gamma_{1}]^{2},[\Gamma_{2}]^{2},$ & $\Gamma_{1}\times\Gamma_{1},\Gamma_{2}\times\Gamma_{2},$	  \\ 
 &  & $B_{g}=\Gamma_{2}$ & $[\Gamma_{3}]^{2},[\Gamma_{4}]^{2}\supset A_{g}$ & $\Gamma_{3}\times\Gamma_{3},\Gamma_{4}\times\Gamma_{4}\supset A_{g}$ \\ 
 & & $A_{u}=\Gamma_{3}$ &  &  \\ 
 & & $B_{u}=\Gamma_{4}$ &  &  \\ \hline
$Y$ $(-\frac{1}{2},\frac{1}{2},0)$ & $C_{2h}$  & $A_{g}=Y_{1}$ & $[Y_{1}]^{2},[Y_{2}]^{2},$ & $Y_{1}\times Y_{1},Y_{2}\times Y_{2},$	  \\ 
 &  & $B_{g}=Y_{2}$ & $[Y_{3}]^{2},[Y_{4}]^{2}\supset A_{g}$ & $Y_{3}\times Y_{3},Y_{4}\times Y_{4}\supset A_{g}$ \\ 
 && $A_{u}=Y_{3}$ &  &  \\ 
& & $B_{u}=Y_{4}$ &  &  \\ \hline 

 $L$ $(-\frac{1}{2},\frac{1}{2},\frac{1}{2})$ & $C_{2h}$  & $A_{g}=L_{1}$ & $[L_{1}]^{2},[L_{2}]^{2},$ & $L_{1}\times L_{1},L_{2}\times L_{2},$	  \\ 
 & & $B_{g}=L_{2}$ & $[L_{3}]^{2},[L_{4}]^{2}\supset A_{g}$ & $L_{3}\times L_{3},L_{4}\times L_{4}\supset A_{g}$ \\ 
& & $A_{u}=L_{3}$ &  &  \\ 
& & $B_{u}=L_{4}$ &  &  \\ \hline 
 
 $Z$ $(0,0,\frac{1}{2})$ & $C_{2h}$ & $A_{g}=Z_{1}$ & $[Z_{1}]^{2},[Z_{2}]^{2},$ & $Z_{1}\times Z_{1},Z_{2}\times Z_{2},$	  \\ 
 &  &$B_{g}=Z_{2}$ & $[Z_{3}]^{2},[Z_{4}]^{2}\supset A_{g}$ & $Z_{3}\times Z_{3},Z_{4}\times Z_{4}\supset A_{g}$ \\ 
 & & $A_{u}=Z_{3}$ &  &  \\ 
& & $B_{u}=Z_{4}$ &  &  \\ \hline 
 
$M$ $(0,\frac{1}{2},\frac{1}{2})$ & $C_{i}$ & $\Gamma_{1,2}\rightarrow M_{1}$ & $[M_{1}]^{2},[M_{2}]^{2}\supset A_{g}$ & $M_{1}\times M_{1},M_{2}\times M_{2}\supset A_{g}$	  \\ 
& & $\Gamma_{3,4}\rightarrow M_{2}$ &  &  \\ \hline 
$N$ $(0,\frac{1}{2},0)$  & $C_{i}$& $\Gamma_{1,2}\rightarrow N_{1}$ & $[N_{1}]^{2},[N_{2}]^{2}\supset A_{g}$ & $N_{1}\times N_{1},N_{2}\times N_{2}\supset A_{g}$	  \\ 
& & $\Gamma_{3,4}\rightarrow N_{2}$ &  &  \\ \hline

$I$ $(\Phi-1,\Phi,\frac{1}{2})$ &$C_{2}$ & $\Gamma_{1,3}\rightarrow I_{1}$ & $[I_{1}]^{2},[I_{2}]^{2}\supset A_{g}$ & $I_{1}\times I_{1},I_{2}\times I_{2}\supset A_{g}$	  \\ 
& & $\Gamma_{2,4}\rightarrow I_{2}$ &  & $I_{1}\times I_{2}\supset B_{g}$  \\ \hline 

$I_{1}$ $(1-\Phi,1-\Phi,\frac{1}{2})$  &$C_{2}$& $\Gamma_{1,3}\rightarrow {I_{1}}_{1}$ & $[{I_{1}}_{1}]^{2},[{I_{1}}_{2}]^{2}\supset A_{g}$ & ${I_{1}}_{1}\times {I_{1}}_{1},{I_{1}}_{2}\times {I_{1}}_{2}\supset A_{g}$	  \\ 
 && $\Gamma_{2,4}\rightarrow {I_{1}}_{2}$ &  & ${I_{1}}_{1}\times {I_{1}}_{2}\supset B_{g}$  \\ \hline 
$X$ $(1-\Psi,1-\Psi,0)$  &$C_{2}$& $\Gamma_{1,3}\rightarrow X_{1}$ & $[X_{1}]^{2},[X_{2}]^{2}\supset A_{g}$ & $X_{1}\times X_{1},X_{2}\times X_{2}\supset A_{g}$	  \\ 
&& $\Gamma_{2,4}\rightarrow X_{2}$ &  & $X_{1}\times X_{2}\supset B_{g}$  \\ \hline 
 
$X_{1}$ $(\Psi-1,\Psi,0)$ &$C_{2}$ & $\Gamma_{1,3}\rightarrow {X_{1}}_{1}$ & $[{X_{1}}_{1}]^{2},[{X_{1}}_{2}]^{2}\supset A_{g}$ & ${X_{1}}_{1}\times {X_{1}}_{1},{X_{1}}_{2}\times {X_{1}}_{2}\supset A_{g}$	  \\ 
& & $\Gamma_{2,4}\rightarrow {X_{1}}_{2}$ &  & ${X_{1}}_{1}\times {X_{1}}_{2}\supset B_{g}$  \\ \hline

$F$ $(\zeta-1,1-\zeta,1-\eta)$  &$C_{s}$ & $\Gamma_{1,4}\rightarrow F_{1}$ & $[F_{1}]^{2},[F_{2}]^{2}\supset A_{g}$ & $F_{1}\times F_{1},F_{2}\times F_{2}\supset A_{g}$	  \\ 
& & $\Gamma_{2,3}\rightarrow F_{2}$ &  & $F_{1}\times F_{2}\supset B_{g}$  \\ \hline 
$F_{1}$ $(-\zeta,\zeta,\eta)$  &$C_{s}$& $\Gamma_{1,4}\rightarrow {F_{1}}_{1}$ & $[{F_{1}}_{1}]^{2},[{F_{1}}_{2}]^{2}\supset A_{g}$ & ${F_{1}}_{1}\times {F_{1}}_{1},{F_{1}}_{2}\times {F_{1}}_{2}\supset A_{g}$	  \\ 
 && $\Gamma_{2,3}\rightarrow {F_{1}}_{2}$ &  & ${F_{1}}_{1}\times {F_{1}}_{2}\supset B_{g}$  \\ \hline

\end{tabular}}
\label{Table: Selection rules_second order}
\end{table}

Owing to momentum conservation, each second-order mode originates from a pair of single-phonon modes located at the same phonon wave vector within the Brillouin zone. With regard to the superposition of single-phonon modes, we discriminate (i) overtones and (ii) combinations, in which two phonons from the (i) same or (ii) different phonon branches at the same critical point combine to produce a two-phonon mode at the two phonons' sum or difference frequency. Combinations or overtones are Raman-active (and thus detectable in second-order Raman measurements) if their reduction to the $\Gamma$ point contains a Raman-active phonon mode. The Raman-active phonons in the monoclinic crystal structure possess $A_{g}$ or $B_{g}$ symmetry, which are detectable using the scattering geometries outlined in \mbox{Table \ref{Table: Selection rules_beta-polymorph}}. $A_{g}$ symmetry modes are, for instance, present when irradiating the (010) plane. As a consequence, only those two-phonon modes whose reduction to the $\Gamma$-point contains an $A_{g}$ representation may occur.   
Four important conclusions can be drawn from \mbox{Table \ref{Table: Selection rules_second order}.}

\begin{enumerate}[I]
    \item Two-phonon modes corresponding to an $A_{g}$ vibrational mode symmetry may originate from all critical points within the Brillouin zone, whereas $B_{g}$ modes are allowed only for critical points exhibiting a $C_{2}$ ($I$, $I_{1}$, $X$ and $X_{1}$) or $C_{s}$ ($F$, $F_{1}$) symmetry.
    \item Overtones always contain the representation $A_{g}$.
    \item Combinations have $A_{g}$ symmetry if phonons originate from dispersion branches of equal irreducible representation.
    \item Combinations of phonons from branches corresponding to different irreducible representations occur only in $B_{g}$ symmetry.
\end{enumerate}

\begin{figure}[!htb]
\centering 
        \includegraphics[width=165mm]{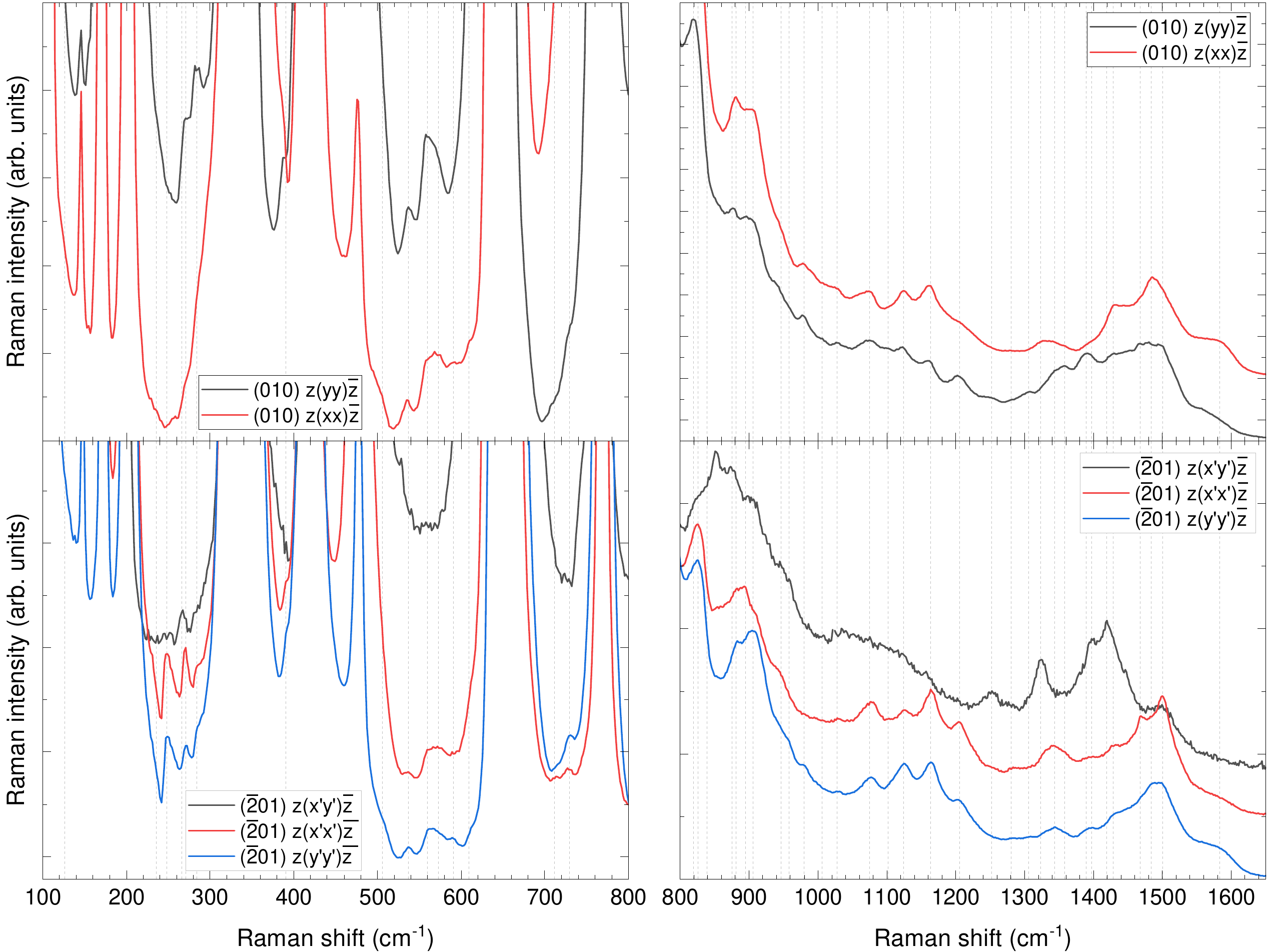}
    \caption[Second-order Raman spectra of monoclinic $\beta$-Ga$_{2}$O$_{3}$]{Second-order Raman spectra of monoclinic $\beta$-Ga$_{2}$O$_{3}$. $A_{g}$-modes are obtained by irradiating the (010) and ($\bar{2}01$) planes in parallel polarization. $B_{g}$-modes are acquired by excitation of the ($\bar{2}01$) plane in crossed polarization. Spectra are vertically-offset and scaled to magnify low-intensity modes and allow for a better comparison between individual spectra. The peak positions of second-order modes are determined by fitting Lorentzian lineshape functions, with the peak positions illustrated by vertical dashed lines. \label{Figure: Second-order Raman spectra}}
\end{figure}

The second-order Raman modes \mbox{(illustrated in Fig. \ref{Figure: Second-order Raman spectra})} of \mbox{$\beta$-Ga$_{2}$O$_{3}$} are obtained in polarized Raman spectroscopic measurements by employing the measurement geometries summarized in \mbox{Table \ref{Table: Selection rules_beta-polymorph},} albeit at longer detection times compared to first-order measurements to resolve modes of low intensities. Peak positions of the two-phonon modes in \mbox{Fig. \ref{Figure: Second-order Raman spectra}} are likewise obtained by fitting Lorentzian lineshape functions, with the peak positions indicated by dashed vertical lines.

An overview of the recorded second-order Raman frequencies in \mbox{Fig. \ref{Figure: Second-order Raman spectra}} and their vibrational symmetries are provided in \mbox{Table \ref{Table: Second-order frequencies, symmetries and assignemts to critical points}.} Vibrational symmetries are deduced from the Raman selection rules summarized in \mbox{Table \ref{Table: Selection rules_beta-polymorph}}, which suggest an $A_{g}$ or $B_{g}$ mode symmetry in parallel polarization or crossed polarization. As pointed out earlier, we discriminate overtones and combinations, to which the recorded second-order Raman peaks are allocated \mbox{(Table \ref{Table: Second-order frequencies, symmetries and assignemts to critical points})} using the following algorithm:

\begin{enumerate}
    \item If a mode occurs in $B_{g}$ symmetry, second-order selection rules \mbox{(Table \ref{Table: Selection rules_second order})} imply that the mode must result from a combinational process occurring at the $I$, $I_{1}$, $X$, $X_{1}$, $F$ or $F_{1}$ point.
    \item $A_{g}$ symmetry modes may originate from either combination or overtone processes at any critical point within the Brillouin zone.
    \item Second-order events are preferably produced by the combination of phonon branches running flat at the critical points within the Brillouin zone, i.e. frequencies characterized by peaks in the PDOS. Thus, if a phonon branch at half the recorded second order Raman frequency has negligible slope or analogously the PDOS exhibits a peak, the two-phonon mode is likely an overtone. Similarly, two distinctive phonons combine if the sum or difference of their frequencies matches a second-order frequency and the respective phonon branches run flat or the PDOS exhibits a peak at either of the two frequencies in question.
    \item The second-order selection rules summarized in \mbox{Table \ref{Table: Selection rules_second order}} must be fulfilled for all considered phonon branch combinations and overtones.
    
\end{enumerate}

In the following, the correlation of a respective second-order mode to its process of creation is illustrated exemplarily for a selection of modes. 
As a first example, we consider the experimental mode recorded at \mbox{1074.3 cm$^{-1}$}. This mode is visible for parallel polarization on the ($\bar{2}01$) and (010) planes, but forbidden for crossed polarization. Thus, first-order Raman selection rules \mbox{(Table \ref{Table: Selection rules_beta-polymorph})} predict that this mode possesses an $A_{g}$ symmetry. The (2) second-order selection rules \mbox{(Table \ref{Table: Selection rules_second order}),} in turn, provide possible critical points from which single phonons may originate to produce second-order events. When exhibiting an $A_{g}$ symmetry, a second-order mode may originate from any critical point and be produced by either an overtone or combinational process on condition that phonon branches exist at the corresponding single-phonon frequencies and these dispersion bands are flat. In order to investigate whether the observed second-order mode is an overtone or combination of phonons we consider the (3) PDOS \mbox{(Fig. \ref{Figure: Phonon dispersion and PDOS}),} which shows an intense peak at about half the recorded second-order frequency. The phonon dispersion branches at the half frequency exist and run flat at the $N$, $X$, $X_{1}$, $M$, $I$ and $I_{1}$ points (cf. last column in \mbox{Table \ref{Table: Second-order frequencies, symmetries and assignemts to critical points}).} Consequently, the observed two-phonon mode likely stems from an optical overtone process, albeit, owing to the rich panoply of phonon branches, combinational modes may still be superimposed.
\newline Similar to the first example, the two-phonon mode recorded at \mbox{572 cm$^{-1}$} (Fig. \ref{Figure: Second-order Raman spectra}) occurs in parallel polarization, thus in $A_{g}$ symmetry. In contrast to the first example, however, this mode's PDOS exhibits no intense peak at half the recorded frequency, ruling out an optical overtone process. Possible combinations require (3) a pair of phonon branches becoming flat at any of the critical points and the sum (or difference) of their frequencies be equal to the observed two-phonon Raman frequency. In combination with (4) the two-phonon Raman selection rules (Table \ref{Table: Selection rules_second order}), possible phonon combinations with phonon dispersion branches running flat are evident at the $N$, $M$, $X$, $X_{1}$, $I$ and $I_{1}$ points. A possible combination characterized by dispersion bands running flat paired with a high PDOS (\mbox{Fig. \ref{Figure: Phonon dispersion and PDOS}}) for the two phonons involved in a sum frequency process is the combination of the lowest-energy acoustic phonon of $N_{1}$ symmetry at \mbox{$\sim 130$ cm$^{-1}$} and the IR-active phonon of the same symmetry predicted at \mbox{$\sim 448$ cm$^{-1}$.} Owing to the high number of phonon branches we do not explicitly list all possible origins of combinations in \mbox{Table \ref{Table: Second-order frequencies, symmetries and assignemts to critical points}}, but limit ourselves to the origins of overtone processes in the following.
\newline 
Whereas the two previous examples of second-order modes occurred in parallel polarization and hence possessed an $A_{g}$ symmetry, modes present solely in crossed polarization on the $(\bar{2}01)$ plane have $B_{g}$ symmetry. An intense $B_{g}$ symmetry second-order mode is located at \mbox{$\sim 851.9$ cm$^{-1}$} \mbox{(cf. Fig. \ref{Figure: Second-order Raman spectra} and Table \ref{Table: Second-order frequencies, symmetries and assignemts to critical points})}. For $B_{g}$ symmetry, (1) second-order selection rules \mbox{(Table \ref{Table: Selection rules_second order})} predict that the second-order mode must stem from a combinational process occurring at the $I$, $I_{1}$, $X$, $X_{1}$, $F$ or $F_{1}$ point. A possible sum frequency combination characterized by a high (3) PDOS of the individual single phonons involves the calculated modes at \mbox{$\sim 154$ cm$^{-1}$} and \mbox{$\sim 676$ cm$^{-1}$.} A possible combination allowed by (4) second-order selection rules \mbox{(Table \ref{Table: Selection rules_second order})} is $X_{1}\times X_{2}$ at the $X-$point.
\newline 
By applying the algorithm explained above we investigate the processes leading to the more than 40 observed second-order Raman peaks \mbox{(Fig. \ref{Figure: Second-order Raman spectra})} and list our findings in \mbox{Table \ref{Table: Second-order frequencies, symmetries and assignemts to critical points}.}
While determining the vibrational symmetries is feasible for the vast majority of two-phonon modes, the two peaks observed at \mbox{819.2 cm$^{-1}$} and \mbox{946.2 cm$^{-1}$} occurred in both $A_{g}$ and $B_{g}$ measurement configurations. We hence list both vibrational symmetries for these two modes in \mbox{Table \ref{Table: Second-order frequencies, symmetries and assignemts to critical points}.} In the case of $B_{g}$ symmetry, the creation process is strictly combinational, whereas in $A_{g}$ symmetry both combinations and overtones may contribute. 
\newline Moreover, the high number of phonon branches within the first Brillouin zone makes the allocation of two-phonon modes to each pair of first-order dispersion branches a challenging endeavour. 
As stated earlier, overtones and combinations may be superimposed provided that the PDOS is intense or the phonon branches run flat at the respective single-phonon frequencies. Nevertheless, the four highest-frequency second-order Raman modes \mbox{($>1468\  \mathrm{cm}^{-1}$)} can, for lack of single phonon branches in the frequency region above \mbox{730 cm$^{-1}$,} clearly be attributed to optical overtone processes. Similarly, acoustic overtones as well as combinations of optical-optical and optical-acoustic phonons are seen in the second-order modes of lower or medium frequencies, also providing access to IR- or acoustic modes.

\definecolor{Gray}{gray}{0.9}
\begin{table}[!htp]
\renewcommand{\arraystretch}{1.0}
\centering
\caption[]{Raman frequencies, vibrational symmetries, processes leading to two-phonon events and origins within the Brillouin zone of \mbox{$\beta$-Ga$_{2}$O$_{3}$} second-order Raman modes observed in \mbox{Fig. \ref{Figure: Second-order Raman spectra}.} The observed peaks are allocated to the two-phonon process and origin of scattering event in the Brillouin zone. Possible origins of overtone processes are listed. $\Gamma$-point overtones are labeled by accounting for the inbound or outbound phonon dispersion branch directions as explained in \mbox{Fig. \ref{Figure: Phonon dispersion and PDOS}}. As an example, a phonon branch possessing a $\Gamma_{1}$ symmetry at the $\Gamma^{Z+}$ point is referred to as $\Gamma^{Z+}_{1}$. }
\begin{tabular}{|c|c|c|c|}
\hline
\rowcolor{Gray} \textbf{Raman shift}& \textbf{Vibrational} & \textbf{Process} & \textbf{Origin in the Brillouin Zone} \\ 
\rowcolor{Gray}\textbf{(cm$^{-1}$)}& \textbf{symmetry} &  & \\ \hline
126.3 & $A_{g}$ & combination  &   \\ \hline
235.4 & $A_{g}$ & optical/acoustic overtone or combination & $[Y_{4}]^{2}$ (optical), $[Z_{4}]^{2}$ (acoustic)\\ \hline
248.0 & $A_{g}$ & combination &  \\ \hline
266.4 & $B_{g}$ & combination &   \\ \hline
271.0 & $A_{g}$ & acoustic overtone or combination  & $[N_{2}]^{2},[X_{2}]^{2},[X_{11}]^{2}$ \\ \hline
284.2 & $A_{g}$ & optical overtone or combination & $[M_{2}]^{2}$ 
\\ \hline
390.6 & $A_{g}$ & optical overtone or combination  & $[Y_{1}]^{2},[M_{1}]^{2},[L_{1}]^{2},[F_{1}]^{2},[Z_{4}]^{2},[F_{11}]^{2},[I_{11}]^{2}$
\\ \hline
506.4 & $A_{g}$ & optical overtone or combination & $[L_{4}]^{2}$ 
\\ \hline
536.7 & $A_{g}$ & optical overtone or combination & $[N_{2}]^{2},[X_{2}]^{2},[M_{1}]^{2},[I_{2}]^{2},[I_{12}]^{2}$
\\ \hline
560.0 & $A_{g}$ & combination &  

\\ \hline
572.7 & $A_{g}$ & combination  &  
\\ \hline
590.5 & $A_{g}$ & combination &  \\ \hline
609.3 & $A_{g}$ & optical overtone or combination & $[N_{1}]^{2},[X_{1}]^{2},[X_{2}]^{2},[X_{11}]^{2},[Y_{2}]^{2},$
\\ 
 &  &  & $[L_{4}]^{2},[F_{2}]^{2},[\Gamma^{Z+}_{3}]^{2},[Z_{2}]^{2},[F_{12}]^{2}$
\\ \hline
711.8 & $A_{g}$ & optical overtone or combination  & $[\Gamma^{Z+}_{1}]^{2},[Y_{3}]^{2},[M_{1}]^{2},[I_{2}]^{2},[L_{2}]^{2},$  \\ 
 &  &   & $[F_{2}]^{2},[Z_{2}]^{2},[F_{12}]^{2},[I_{12}]^{2}$   \\ \hline
728.1 & $A_{g}$ & combination &  \\ \hline
819.2 & $A_{g}$ or $B_{g}$ & optical overtone or combination & $[N_{2}]^{2},[M_{1}]^{2},[M_{2}]^{2},[I_{1}]^{2},[I_{2}]^{2},[I_{12}]^{2}$ \\ \hline
826.2 & $A_{g}$ & optical overtone or combination & $[N_{2}]^{2},[M_{1}]^{2},[M_{2}]^{2},[I_{1}]^{2},[I_{2}]^{2},[I_{12}]^{2}$  \\ \hline
851.9 & $B_{g}$ & combination &   \\ \hline
874.9 & $B_{g}$ & combination &  \\ \hline
881.0 & $A_{g}$ & optical overtone &  $[N_{1}]^{2},[X_{1}]^{2},[X_{2}]^{2},[M_{2}]^{2},[I_{1}]^{2},[I_{11}]^{2}$  \\ \hline
895.8 & $A_{g}$  & optical overtone & $[N_{1}]^{2},[X_{2}]^{2},[Y_{2}]^{2},[M_{2}]^{2},[I_{1}]^{2},[I_{11}]^{2}$ \\ \hline
907.9 & $A_{g}$ & optical overtone or combination & $[N_{2}]^{2},[X_{1}]^{2},[X_{11}]^{2},[\Gamma_{3}^{X-}]^{2},[Y_{2}]^{2},$  \\ 
 &  &  & $[M_{2}]^{2},[I_{1}]^{2},[L_{2}]^{2},[F_{2}]^{2},[Z_{3}]^{2},[F_{12}]^{2}$  \\ \hline
946.2 & $A_{g}$ or $B_{g}$ & optical overtone or combination & $[\Gamma_{2}^{Z+}]^{2},[Z_{1}]^{2}$   \\ \hline
957.5 & $A_{g}$  & optical overtone or combination & $[\Gamma_{1}^{M+}]^{2},[\Gamma_{2}^{Z+}]^{2},[F_{11}]^{2}$  \\ \hline
979.5 & $A_{g}$ & combination &  \\ \hline
1005.8 & $A_{g}$ & combination &  \\ \hline
1027.6 & $A_{g}$ & combination &   \\ \hline
1074.3 & $A_{g}$ & optical overtone or combination &  $[N_{1}]^{2},[X_{2}]^{2},[X_{12}]^{2},[M_{1}]^{2},[I_{2}]^{2},[I_{12}]^{2}$ \\ \hline
1102.0 & $A_{g}$ & combination &  \\ \hline
1124.3 & $A_{g}$ & optical overtone  & $[N_{2}]^{2},[X_{1}]^{2},[M_{2}]^{2},[I_{1}]^{2},[I_{11}]^{2}$\\ \hline
1163.4 & $A_{g}$ & combination &  \\ \hline
1205.0 & $A_{g}$ & optical overtone or combination & $[N_{1}]^{2},[M_{2}]^{2},[I_{2}]^{2}$  \\ \hline
1251.8 & $B_{g}$ & combination &  \\ \hline
1280.3 & $A_{g}$  & overtone or combination & $[X_{1}]^{2},[M_{2}]^{2},[I_{1}]^{2},[Z_{4}]^{2},[I_{11}]^{2}$ \\ \hline
1305.4 & $A_{g}$ & optical overtone  & $[\Gamma_{1}^{Z+}]^{2},[M_{1}]^{2},[I_{2}]^{2},[F_{2}]^{2},$  \\ 
 &  &   & $[F_{12}]^{2},[I_{12}]^{2},[Z_{3}]^{2},[Z_{4}]^{2}$  \\ \hline
1323.6 & $B_{g}$ & combination &   \\ \hline
1343.5 & $A_{g}$ & optical overtone or combination &  $[N_{2}]^{2},[X_{2}]^{2},[X_{12}]^{2}$  \\ \hline
1357.7 & $A_{g}$ & optical overtone & $[N_{2}]^{2},[X_{2}]^{2},[X_{12}]^{2}$  \\ \hline
1389.0 & $A_{g}$ & optical overtone or combination &  $[N_{1}]^{2},[X_{1}]^{2},[I_{1}]^{2},[M_{2}]^{2},[I_{2}]^{2}$  \\ \hline
1397.0 & $B_{g}$ & combination  &   \\ \hline
1418.8 & $B_{g}$ & combination  & \\ \hline
1428.2 & $A_{g}$ & optical overtone or combination &  $[M_{1}]^{2},[I_{1}]^{2},[I_{11}]^{2}$  \\ \hline
1468.0 & $A_{g}$ & optical overtone & $[N_{1}]^{2},[N_{2}]^{2},[X_{1}]^{2},[X_{2}]^{2},[X_{11}]^{2},$ \\ 
 &  &  & $[X_{12}]^{2},[Y_{1}]^{2},[Y_{4}]^{2},[I_{12}]^{2}$ \\ \hline
1483.9 & $A_{g}$ & optical overtone & $[N_{1}]^{2},[X_{1}]^{2},[X_{11}]^{2},[Y_{1}]^{2}$ \\ \hline
1500.5 & $A_{g}$ & optical overtone & $[\Gamma_{1}^{X-}]^{2},[\Gamma_{4}^{M+}]^{2}$ \\ \hline
1582.8 & $A_{g}$ & optical overtone & $[L_{1}]^{2},[Z_{4}]^{2}$  \\ \hline
\end{tabular}
\label{Table: Second-order frequencies, symmetries and assignemts to critical points}
\end{table}

\section{Conclusions}

We employed a combination of polarized micro-Raman spectroscopy and density functional perturbation theory calculations to investigate the first- and second-order Raman spectra of $\beta$-Ga$_{2}$O$_{3}$ grown by the Czochralski method. First, the frequencies of all 15 Raman-active first-order phonons were obtained, with the frequency difference between the closely-matching $A^{7}_{g}$ and $B^{4}_{g}$ modes being determined to \mbox{1.0 cm$^{-1}$}. Secondly, the recorded second-order Raman spectra were analyzed in conjunction with the calculated phonon dispersion, phonon density of states and the derived group-theoretical selection rules for two-phonon scattering. In detail, we identified more than 40 second-order modes and attributed their creation to phonon overtone or combination processes. While modes of $B_{g}$ symmetry always originate from combinations, $A_{g}$ modes may be both overtones and combinations.  
This study enables to study and identify scattering processes of phonons from across the Brillouin zone of monoclinic \mbox{$\beta$-Ga$_{2}$O$_{3}$} and provides access to both IR- and Raman-active phonons via Raman spectroscopy.

\section*{Acknowledgements}
We acknowledge funding by the Deutsche Forschungsgemeinschaft (DFG, German Research Foundation) - project number 446185170. This work was performed in parts in the framework of GraFOx, a Leibniz-ScienceCampus partially funded by the Leibniz association. Computational resources used for the calculations were provided by the HPC of the Regional Computer Centre Erlangen (RRZE). Dr. Klaus Irmscher (Leibniz-Institut für Kristallzüchtung) is appreciated for critical reading of the present paper.

\newpage
\setcounter{page}{1}
\setcounter{figure}{0}
\setcounter{section}{0}
\setcounter{table}{0}

\renewcommand{\baselinestretch}{1.33} 
\renewcommand{\thepage}{S\arabic{page}}
\renewcommand{\thetable}{S\arabic{table}}
\renewcommand{\thefigure}{S\arabic{figure}}
\renewcommand{\theequation}{S\arabic{equation}}
\renewcommand{\thesection}{S\arabic{section}}

\section{\Large{Supplementary Material}}

\subsection{Phonon dispersion and density of state}

\begin{figure}[!htb]
\centering 
        \includegraphics[width=165mm]{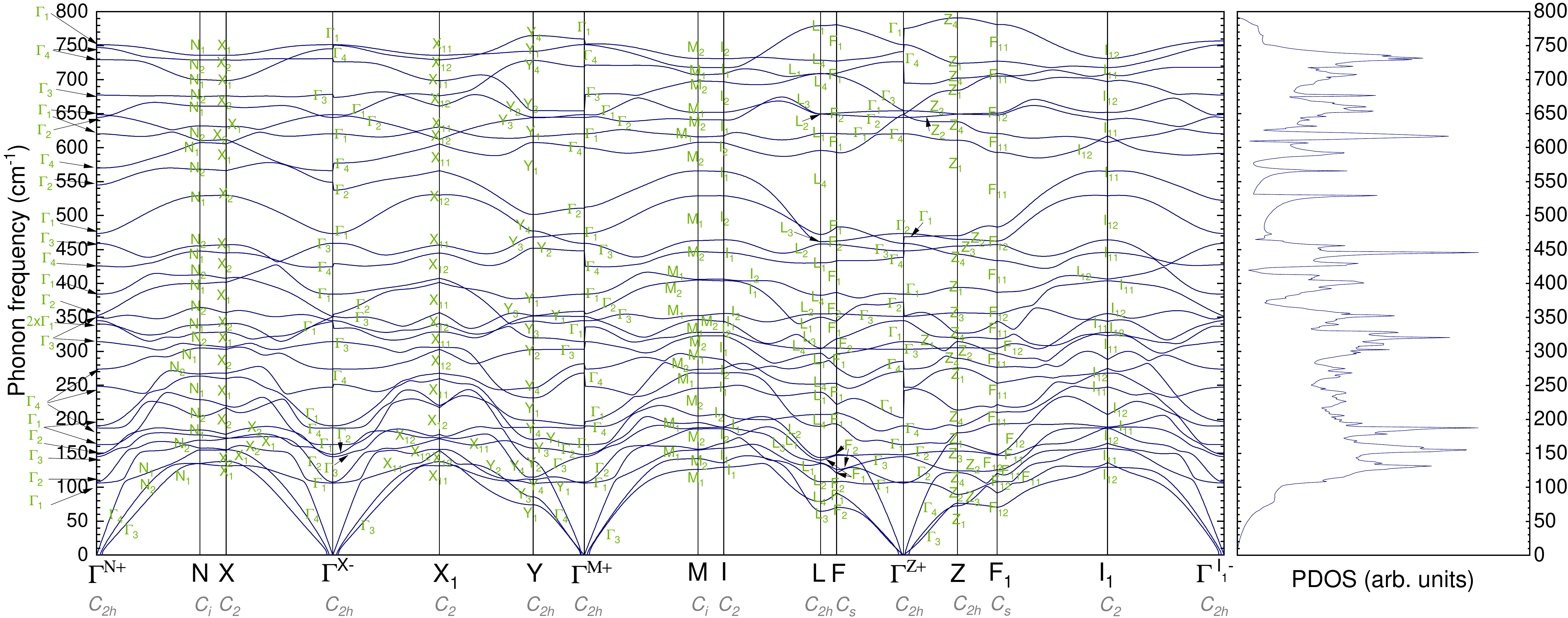}
    \caption[Calculated phonon dispersion and phonon density of states of monoclinic $\beta$-Ga$_{2}$O$_{3}$]{Calculated phonon dispersion (left) and phonon density of states (right) of monoclinic \mbox{$\beta$-Ga$_{2}$O$_{3}$.} The critical points in the monoclinic unit cell along with their corresponding site symmetry groups are illustrated at the bottom of the graph. Individual $\Gamma$-points are distinguished by introducing supercripts to label each $\Gamma$-point by the inbound (-) or outbound (+) phonon branch direction, e.g. $\Gamma^{X-}$ or $\Gamma^{X_{1}+}$ as equivalent labels for the $\Gamma$-point between the $X$ and $X_{1}$ points. "PDOS" denotes phonon density of states.  \label{Figure: Phonon dispersion and PDOS_mit Labels}}
\end{figure}

Fig. \ref{Figure: Phonon dispersion and PDOS_mit Labels} illustrates the calculated dispersion curves and PDOS, indicating the symmetries of all phonon branches within the Brillouin zone. 

\subsection{Lattice parameters used in the current study and previous works}

Table \ref{tab:latconst} lists the lattice parameters employed in the DFPT calculations of this work. The utilized values are consistent with previous experimental and theoretical studies.

\begin{table}[H]
\renewcommand{\arraystretch}{1.3}
\caption{ Lattice constants $a$, $b$, $c$ and the monoclinic angle $\beta$ obtained from DFT-LDA calculations in this work compared to experimental and theoretical data. The exchange-correlation functional applied in each theoretical work is specified. }
\begin{tabular}{ c | c | c | c | c}
\hline\hline
        & a (\si{\angstrom}) & b (\si{\angstrom}) & c (\si{\angstrom}) & $\beta$ ($^{\circ}$)\\\hline
LDA$^{\ddagger}$& 12.155 & 3.022  & 5.758 & 103.71\\
LDA\cite{Janzen2021}& 12.154 & 3.022  & 5.759 & 103.31\\
Experiment\cite{Ahman1996}   & 12.21  & 3.04   & 5.80  & 103.83\\
AM05\cite{Furthmueller}    & 12.29  & 3.05   & 5.81  & 103.77\\
PBE\cite{Yoshioka}    & 12.44  & 3.08   & 5.88  & 103.71\\
PBE\cite{Liu}    & 12.31  & 3.08   & 5.89  & 103.9\\
LDA\cite{Yamaguchi2004}    & 12.23  & 3.04   & 5.80  & 103.7\\
\hline\hline
\end{tabular}
\label{tab:latconst}
\footnoterule{$^{\ddagger}$: This work }
\end{table}

\clearpage
\newpage
\setcounter{page}{1}
\pagenumbering{roman}
\bibliographystyle{unsrt}


\end{document}